\documentstyle[ltwol,epsfig]{article}

\def\NIMA{{\em Nucl. Instrum. Methods} A}
\def\NPB{{\em Nucl. Phys.} }
\def\PLB{{\em Phys. Lett.} }

\def\be{\begin{equation}}
\def\ee{\end{equation}}
\def\bea{\begin{eqnarray}}
\def\eea{\end{eqnarray}}

\begin{document}

\title{CHARMLESS $B$ DECAYS INTO THREE CHARGED TRACK FINAL STATES}

\author{A. GARMASH}

\address{ Budker Institute of Nuclear Physics, Novosibirsk, 630090, Russia \\
                                       and\\
KEK, High Energy Accelerator Research Organization, Oho 1-1, Tsukuba, Ibaraki, Japan\\
                          E-mail: garmash@bmail.kek.jp
\vspace*{0.1cm} \\
                          For the Belle Collaboration}


\twocolumn[\maketitle\abstracts{ 
    Using a data sample of 10.5 fb$^{-1}$ collected by the Belle detector, 
three--body charmless decays $B^+\to K^+h^+h^-$ have been studied.
The following branching fractions have been obtained:
${\cal{B}}(B^+\to K^+\pi^-\pi^+) = (64.8\pm10.0\pm7.0)\times10^{-6}$ and
${\cal{B}}(B^+\to K^+K^-K^+)     = (36.5\pm6.1\pm5.5)\times10^{-6}$. 
The upper limits for other combinations of charged kaons and pions have 
been placed. Analysis of the intermediate two-body states gives evidence 
for production of scalar resonances in charmless $B$ decays.
}]

\section{Introduction}
   Until recently charmless B decays were mainly studied via their
two-body decay modes  because of the large combinatorial background
in multibody channels.
In this work we attempt to  study $B^+ \to K^+h^+h^-$ decays
($h$ stands for a charged pion or kaon) without any assumption about
the intermediate hadronic resonances. 
Reference to the charge conjugate states is implicit 
throughout this paper unless explicitly stated otherwise.

Analysis is based on the data set collected by the Belle detector~\cite{Belle} 
at KEKB~\cite{KEKB}, the asymmetric
B-factory at KEK. It consists of 10.5 fb$^{-1}$ taken at the $\Upsilon(4S)$
resonance and 0.6 fb$^{-1}$ taken below the $B\bar{B}$ production threshold 
for continuum studies.

\section{Event selection}
Charged tracks are required to satisfy a set of track quality cuts
based on the the average hit residual and the impact parameters in both the 
$r-\phi$ and $r-z$ planes. We require that the transverse track momenta be
greater than 100 MeV/$c$ to reduce low momentum combinatorial background.

   The candidate events are identified by using the beam--constrained mass 
$M_{bc} = \sqrt{(\sqrt{s}/2)^2 - P_B^{*2} }$
and the calculated energy difference $\Delta E = E_B^* - \sqrt{s}/2$ 
where $E_B^*$ and $P_B^*$ are the energy and 3--momentum of the $B$ 
candidate in the $\Upsilon (4S)$ rest frame.
We define the $B$ signal region as: $5.271<M_{bc}<5.289$ GeV/$c^2$;~$|\Delta E|<40$~MeV.

   The most important issues of this analysis are efficient kaon 
identification over the whole momentum region and the suppression of the large
combinatorial background which is dominated by $q\bar{q}$ continuum
production. 

   Since two $B$ mesons produced from the $\Upsilon (4S)$ decay are 
nearly at rest in the CMS frame, the angles of the decay products
of two $B$'s are uncorrelated and the event looks spherical. In contrast,
hadrons from continuum $q\bar{q}$ events tend to exhibit a two-jet 
structure.

   We calculate  the angle, $\theta_{Thr}$, between the thrust axis of 
the $B$ candidate and that of the rest of the event. 
The distribution of $|\cos(\theta_{Thr})|$ is strongly
peaked near 1.0 for $q\bar{q}$ events while it is nearly flat for
$B\bar{B}$ events.
We require $|\cos(\theta_{Thr})|<0.80$ for all modes under consideration.

   We also form a Fisher discriminant~\cite{Fisher} with the momentum scalar 
sum of charged particles and photons in nine cones of increasing polar angle
around the thrust axis of the $B$ candidate and the angle of the thrust axis 
of the candidate with respect to the beam axis. We combine the 
Fisher discriminant and $\cos(\theta_{B})$, 
where $\theta_{B}$ is the angle between the $B$ candidate momentum 
and the beam axis, in a single variable  by taking a product 
of the corresponding
probability density functions and impose a cut on the likelihood ratio.

   Separation of kaons and pions is accomplished by combining
the responses of the ACC and the TOF with
$dE/dx$ measurements in the CDC.  
The combined response of the three systems 
provides high quality $K/\pi$ separation
in the laboratory momentum range up to 3.5 GeV/$c$.
At large momenta
only the ACC and $dE/dx$ are used since here
TOF provides no significant separation of kaons and pions.  

\section{Analysis}
In this analysis at least one charged track was required to be positively
identified as a kaon. Then we considered all possible combinations of 
charged kaons and pions. The analysis of $K^+\pi^+\pi^-$ and 
$K^+K^+K^-$ final states is described in detail in two following
subsections. The results on the study of other combinations
are summarized in Table~\ref{tab:result1}.

\begin{figure}[t]
\center
  \begin{tabular}[h]{l}
    \hspace*{-0.1cm}\epsfig{figure=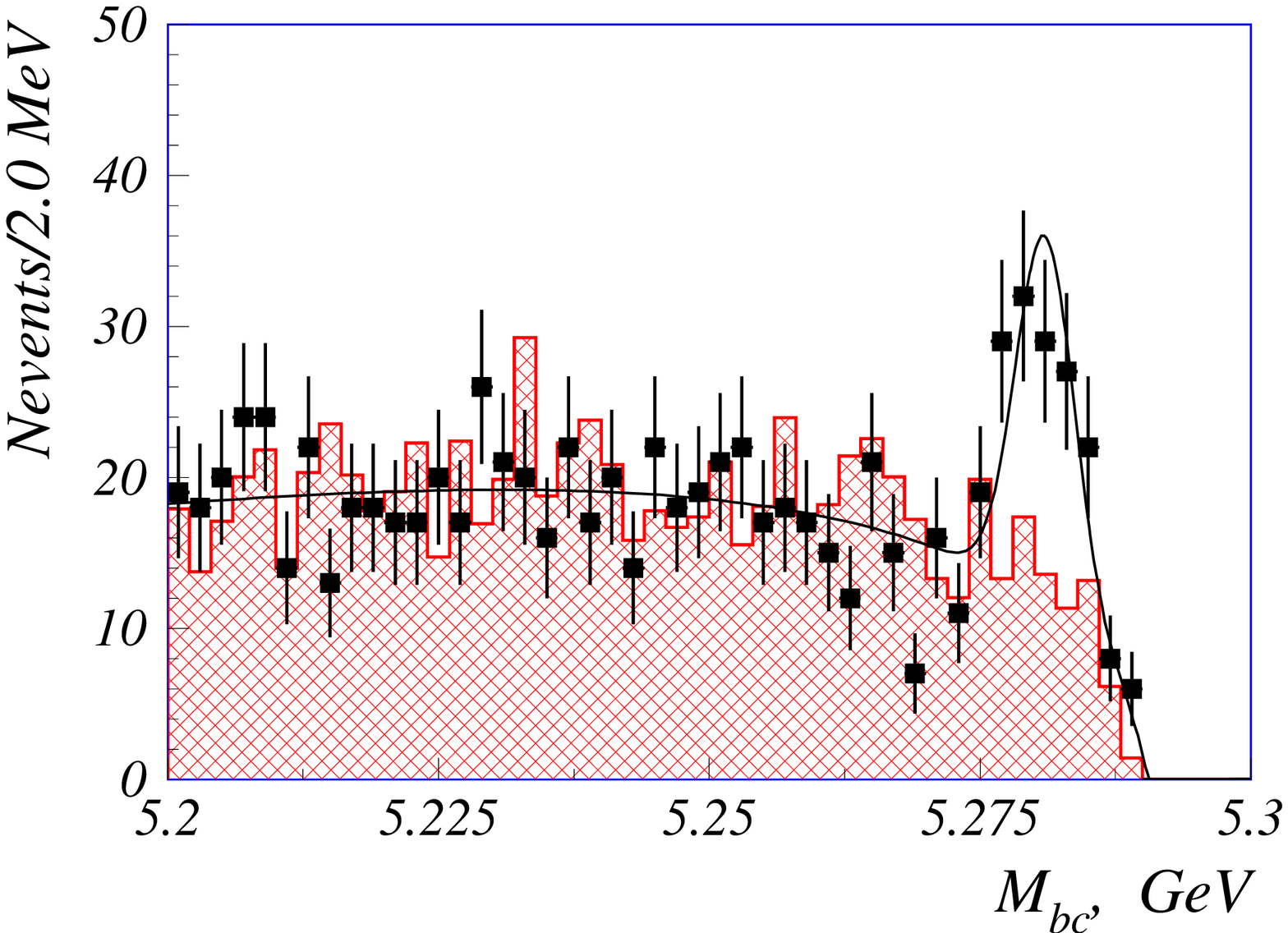,height=1.2in,width=3.6in}\\
    \hspace*{-0.1cm}\epsfig{figure=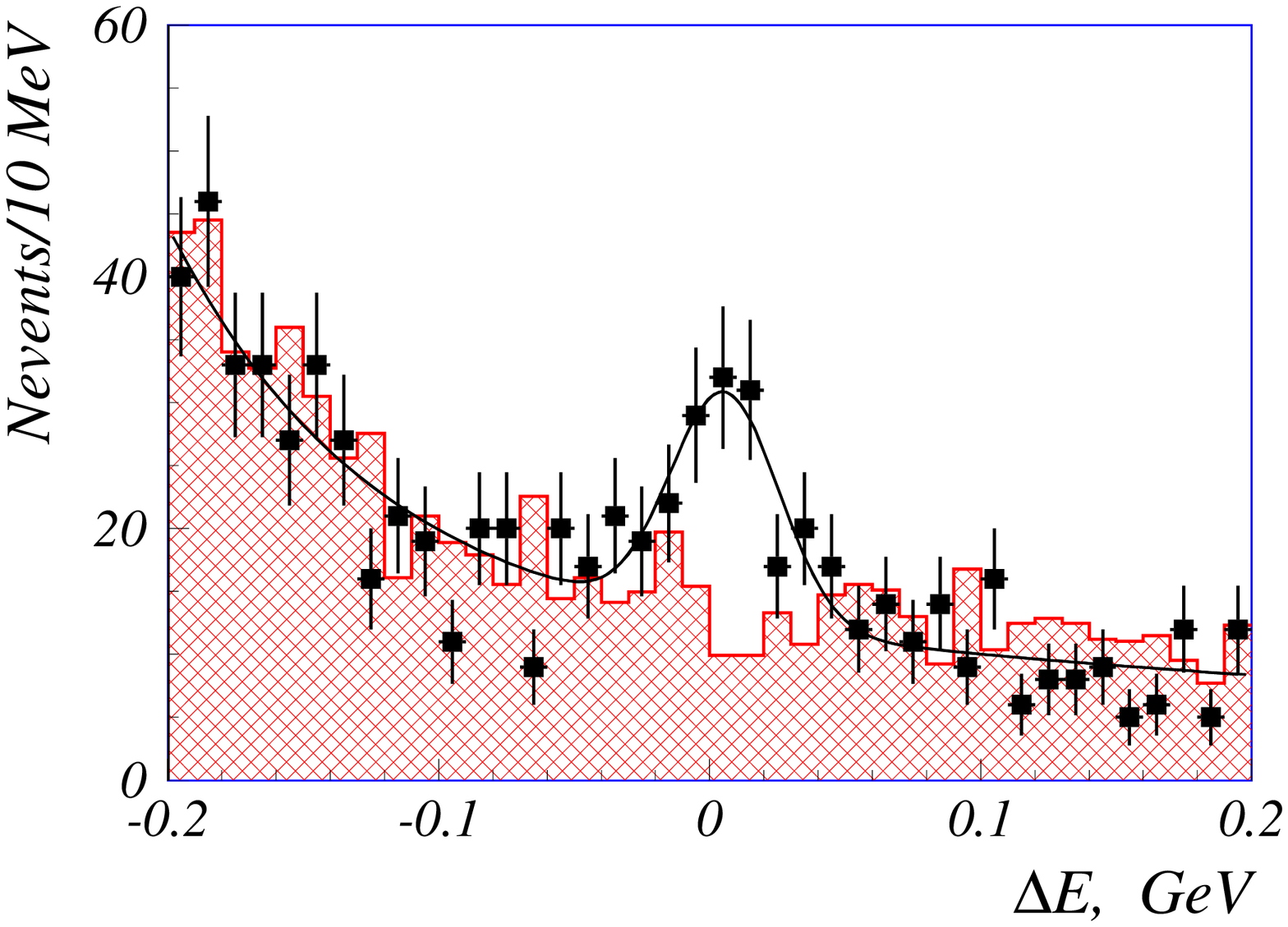,height=1.2in,width=3.6in}
  \end{tabular}
  \caption{The $M_{bc}$ (top) and $\Delta E$ (bottom) distributions for selected
           $B^+\to K^+\pi^+\pi^-$ candidates. Points are data and histograms are 
           Monte Carlo expectation for background. Curves are fit to the data.}
  \label{fig:mbde_kpp}
\end{figure}
\vspace*{-1.8pt}   
\subsection{$B^+\to K^+\pi^+\pi^-$}
To study this final state we require one track to be positively
identified as a kaon and two tracks consistent with the pion hypothesis.
The contributions from
$B^+ \to \bar{D}^0\pi^+$ where $\bar{D}^0 \to K^+\pi^-$ and
$B^+ \to J/\psi(\psi') K^+$ where $J/\psi(\psi') \to \mu^+\mu^-$
submodes were excluded from the analysis by imposing cuts on the invariant 
masses of two intermediate particles:
  $|M(K^+\pi^-)-1.865| > 0.100$ GeV/$c^2$;
  $|M(h^+h^-)-3.097|   > 0.070$ GeV/$c^2$;
  $|M(h^+h^-)-3.686|   > 0.050$ GeV/$c^2$
where $h^+$ and $h^-$ are pion candidates.
The latter two submodes contribute due to the muon-pion misidentification 
and in these cases we use the muon mass for the $M(h^+h^-)$ calculation.

   The $M_{bc}$ and $\Delta E$  distributions
after excluding these signals are presented in Fig.~\ref{fig:mbde_kpp}.
A significant enhancement in the $B$ signal region can be seen in both
distributions.

\begin{figure}
\center
  \begin{tabular}[h]{l}
    \hspace*{-0.1cm}\epsfig{figure=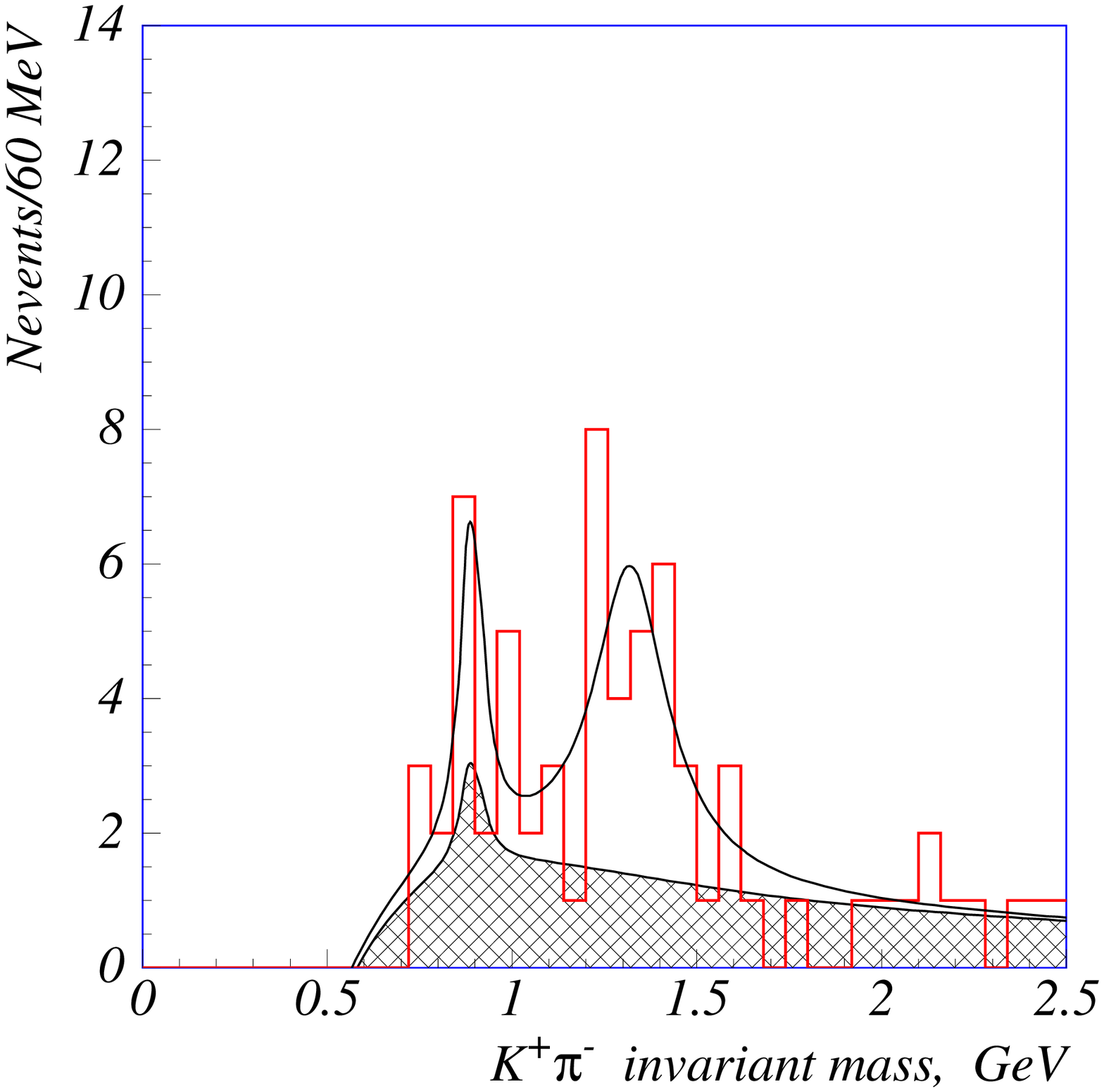,height=1.2in,width=3.6in}\\
    \hspace*{-0.1cm}\epsfig{figure=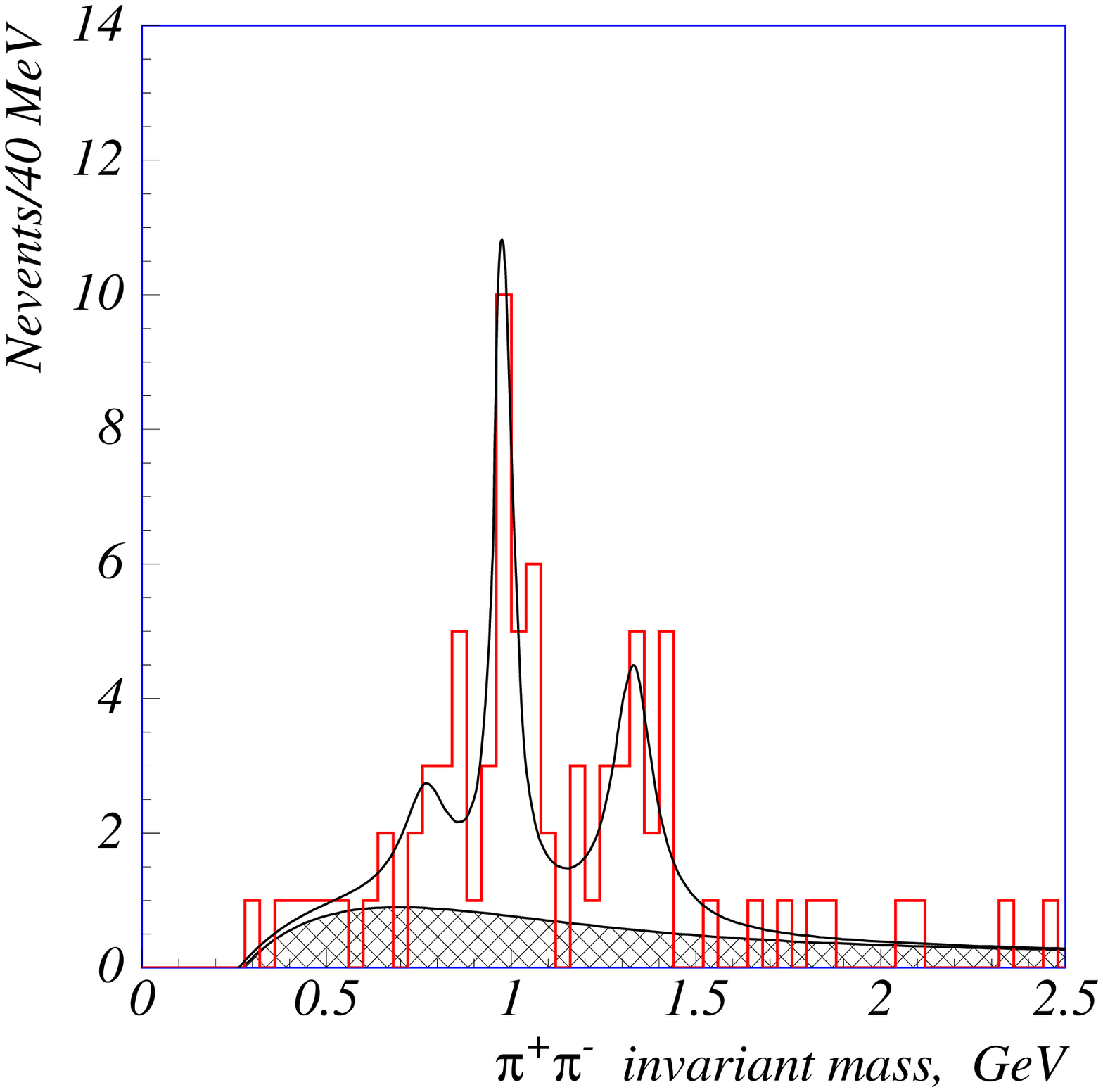,height=1.2in,width=3.6in}
  \end{tabular}
  \caption{The $K^+\pi^-$ (top) and $\pi^+\pi^-$ (bottom) invariant mass spectra for 
           candidates from the $B$ signal box. Histograms are data and curves are 
           fitting results. The hatched regions are the background estimates from 
           sidebands.}
  \label{fig:kpmass}
\end{figure}

  To determine the intermediate states which contribute to the observed signal, we plot 
the $K^+\pi^-$ and $\pi^+\pi^-$ invariant mass spectra as shown in Fig.~\ref{fig:kpmass}.
The dashed regions in Fig.~\ref{fig:kpmass} show the background spectra determined from
$M_{bc}$ and $\Delta E$ sidebands. We fit the signal in the $K^+\pi^-$ 
invariant mass spectrum to the
non-coherent sum of two relativistic Breit-Wigner functions. The parameters of the first
Breit-Wigner were fixed to be equal to those of the $K^{*o}(892)$ meson and the parameters
of the second one referred to as $K^{*o}_X(1400)$ were free during the fit. 
The signal in the $\pi^+\pi^-$ invariant mass spectrum is
fitted to the non-coherent sum of three relativistic Breit-Wigner functions.
The parameters of the first and second Breit-Wigners were fixed to be equal to those 
of $\rho^o(770)$ and $f_o(980)$ mesons respectively and the parameters of the
third one referred to as $f_X(1300)$ were free during the fit. Results of the fit are 
summarized in Table~\ref{tab:kkfit}.

\vspace*{-1.8pt}   
\subsection{$B^+\to K^+K^+K^-$}
We select $B$ candidates formed from three charged tracks 
identified as kaons. The contribution from the Cabibbo suppressed
$B^+\to D^oK^+$ decay where $D^o\to K^+K^-$, is
excluded from the analysis by imposing cuts on the $K^+K^-$ invariant mass:
$|M(K^+K^-)-1.865| > 0.025$ GeV/$c^2$.
To suppress the background from $\pi$--$K$ misidentification we
reject all candidates if the invariant mass of any two oppositely charged
tracks is consistent within $2\sigma$ with the $D\to K\pi$ hypothesis 
with the kaon hypothesis assigned to be either of the two tracks. No particle
identification information is used in this veto.

\begin{figure}
\center
  \begin{tabular}[t]{l}
    \hspace*{-0.1cm}\epsfig{figure=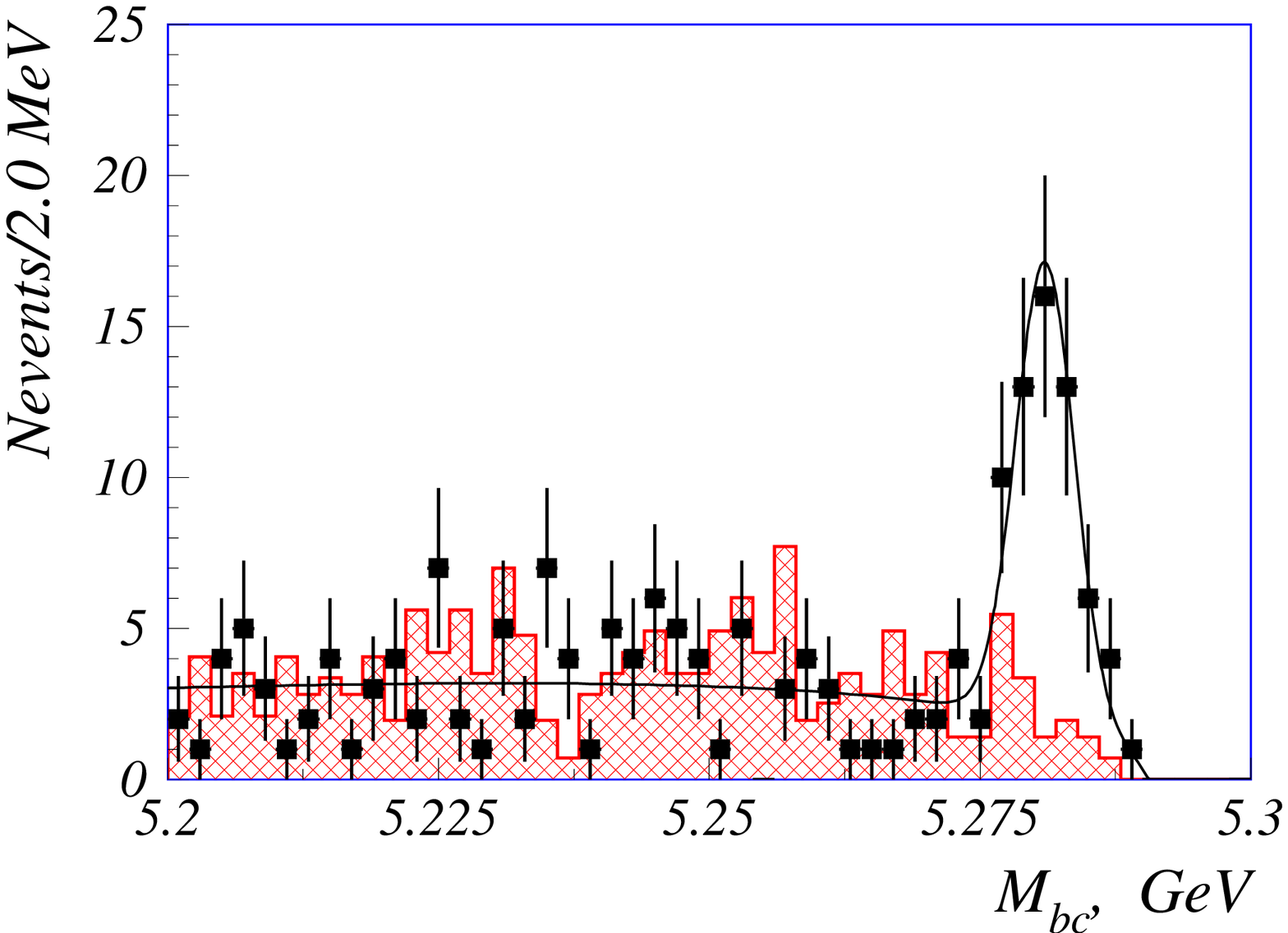,height=1.2in,width=3.6in}\\
    \hspace*{-0.1cm}\epsfig{figure=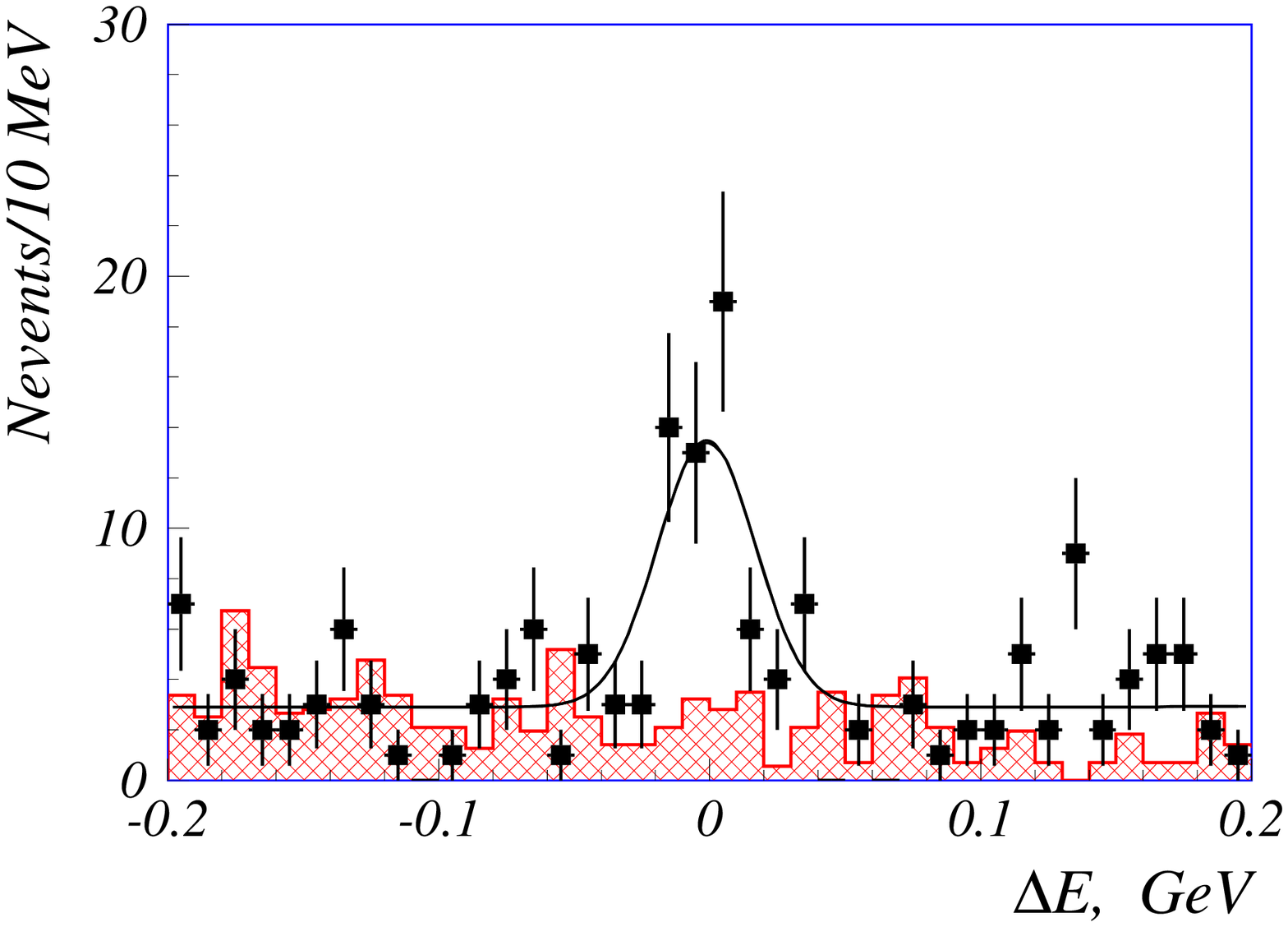,height=1.2in,width=3.6in}
  \end{tabular}
  \caption{The $M_{bc}$ (top) and $\Delta E$ (bottom) distributions for selected
           $B^+\to K^+K^+K^-$ candidates. Points are data and histograms are 
           Monte Carlo expectation for background. Curves are fit to the data.}
  \label{fig:mbde_kkk}
\end{figure}

   The resulting $M_{bc}$ and $\Delta E$ distributions are 
presented in Fig.~\ref{fig:mbde_kkk}.
A large enhancement in the $B$ signal region can be seen in both
distributions.

\begin{figure}[t]
\center
  \hspace*{-0.1cm}\epsfig{figure=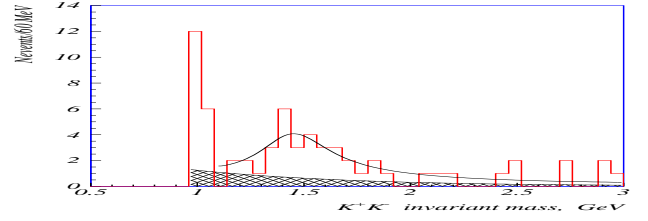,height=1.2in,width=3.6in}
  \caption{The $K^+K^-$ invariant mass spectrum for candidates from the $B$ 
           signal box. The hatched region is the background estimates from 
           sidebands.}
  \label{fig:kkmass}
\end{figure}

  To determine the intermediate states which contribute to the observed signal, 
we examine the $K^+K^-$ invariant mass spectrum which is shown in Fig.~\ref{fig:kkmass}.
The dashed region in Fig.~\ref{fig:kkmass} shows the background spectrum determined 
from $M_{bc}$ and $\Delta E$ sidebands.

   The $K^+K^-$ invariant mass spectrum is characterized by two distinct 
structures. The first one around 1.0~GeV/$c^2$ is very narrow and corresponds to the
$\phi(1020)$ meson. The other one around 1.5~GeV/$c^2$ which we refer to as 
$f_X(1500)$ is very broad. 
We fit the signal in the $K^+K^-$ invariant mass spectrum to the
relativistic Breit-Wigner function. The $K^+K^-$ invariant mass region
correspondig to the $\phi(1020)$ meson was excluded from the fit.
The results of the fit are summarized in Table~\ref{tab:kkfit}.

\section{Results}
The final signal yield for each submode was extracted from the fit to the beam 
constrained mass distributions after cuts on the invariant mass of two 
intermediate particles are applied. We fit the $M_{bc}$ distributions to
the sum of a Gaussian function for signal and ARGUS function~\cite{ARGUS} 
for background. We assume that yield in each bin of $K^+\pi^-$/$\pi^+\pi^-$
is entirely due to a single resonance.

   To reduce the systematic error in the branching fractions we 
normalize our results to the $B^+\to \bar{D}^0\pi$, $\bar{D}^0\to K^+\pi^-$ signal. 
Similarly, because of poor knowledge of the absolute branching fractions
and the uncertainty in the interpretation of some
of intermediate resonances, we calculate only the
products of the branching fractions rather than their absolute values:
$$
  {\cal{B}}_{B^+\to Xh^+}\times{\cal{B}}_{X\to h^+h^-} = 
   \frac{N_X}{N_{D\pi}}\times{\cal{B}}_{B\to D\pi}
   \times{\cal{B}}_{D\to K\pi}\times\frac{1}{\delta}
$$
where $X$ denotes a $h^+h^-$ resonance state, 
$N_X$ and $N_{D\pi}$ are the number of observed events for a mode under study and for
the reference process respectively,
${\cal{B}}_{B\to D\pi}$ and ${\cal{B}}_{D\to K\pi}$ are branching fractions for 
$B^+\to \bar{D}^0\pi$ and $\bar{D}^0\to K^+\pi^-$ respectively~\cite{PDG} and
$\delta$ is the overall correction factor.

\section{Discussion \& Conclusions}
The high quality of $\pi/K$ separation allowed us to measure for the 
first time the branching ratios of the three-body decays 
$B^+ \to K^+\pi^+\pi^-$ and $B^+ \to K^+K^+K^-$ without any assumptions
about particular intermediate mechanisms. CLEO~\cite{berg96b} and
BaBar~\cite{babar} have previously placed upper limits on non-resonance 
three-body decays.
The reported numbers for
$B^+ \to K^+\pi^+\pi^-$ (CLEO: $<28 \times 10^{-6}$, BaBar: $<66 \times 10^{-6}$)
and $B^+ \to K^+K^+K^-$ (CLEO: $<38 \times 10^{-6}$)
are considerably lower than those presented in this paper.
Comparison of the applied selection criteria 
shows that CLEO and BaBar restricted their analysis to the
region of the invariant masses above 2~GeV/$c^2$ for any pair of the
particles. Assuming the phase space distribution of
the invariant  masses, they obtained the limits quoted above.
Similar analysis in our case gives consistent  results.

   The upper limits obtained for the $K^-\pi^+\pi^+$, $K^+K^-\pi^+$ and $K^+K^+\pi^-$
modes are considerably better than previous results by 
CLEO~\cite{berg96b} and OPAL~\cite{OPAL}. A search for the $K^+K^+\pi^-$ mode 
is of particular interest since in some extensions of the Standard Model 
this branching fraction is predicted to be significantly enhanced.

\begin{table}[t]
\begin{center}
\caption{Results of the fit to the $K^+\pi^-$, $\pi^+\pi^-$ and $K^+K^-$ invariant mass spectra.}
\label{tab:kkfit}
\vspace{0.2cm}
\begin{tabular}{|c|c|c|c|} 
\hline 
\raisebox{0pt}[12pt][6pt]{State}         &
\raisebox{0pt}[12pt][6pt]{M, GeV/$c^2$}  &
\raisebox{0pt}[12pt][6pt]{$\Gamma$, GeV} &
\raisebox{0pt}[12pt][6pt]{Yield} \\
\hline
\raisebox{0pt}[12pt][6pt]{$K^{*o}(892)$} &
\raisebox{0pt}[12pt][6pt]{0.896}         &
\raisebox{0pt}[12pt][6pt]{0.051}         &
\raisebox{0pt}[12pt][6pt]{$8.7\pm4.7$} \\
\hline
\raisebox{0pt}[12pt][6pt]{$K^{*o}_X(1400)$} & 
\raisebox{0pt}[12pt][6pt]{$1.31\pm0.05$}    & 
\raisebox{0pt}[12pt][6pt]{$0.23^{+0.10}_{-0.07}$} & 
\raisebox{0pt}[12pt][6pt]{$28.3\pm7.4$} \\ 
\hline
\raisebox{0pt}[12pt][6pt]{$\rho^o(770)$} &
\raisebox{0pt}[12pt][6pt]{0.769}         &
\raisebox{0pt}[12pt][6pt]{0.151}         &
\raisebox{0pt}[12pt][6pt]{$10.2\pm5.8$} \\
\hline
\raisebox{0pt}[12pt][6pt]{$f_o(980)$}    &
\raisebox{0pt}[12pt][6pt]{0.980}         &
\raisebox{0pt}[12pt][6pt]{0.060}         &
\raisebox{0pt}[12pt][6pt]{$22.8\pm4.9$} \\
\hline
\raisebox{0pt}[12pt][6pt]{$f_X(1300)$}   &
\raisebox{0pt}[12pt][6pt]{$1.33\pm0.03$} &
\raisebox{0pt}[12pt][6pt]{$0.13^{+0.07}_{-0.04}$} &
\raisebox{0pt}[12pt][6pt]{$17.7\pm6.6$} \\
\hline
\raisebox{0pt}[12pt][6pt]{$\phi(1020)$}  &
\raisebox{0pt}[12pt][6pt]{1019.4}        &
\raisebox{0pt}[12pt][6pt]{-}             &
\raisebox{0pt}[12pt][6pt]{$14.1\pm4.6$} \\
\hline
\raisebox{0pt}[12pt][6pt]{$f_X(1500)$}   &
\raisebox{0pt}[12pt][6pt]{$1.43\pm0.08$} &
\raisebox{0pt}[12pt][6pt]{$0.41^{+0.18}_{-0.18}$} &
\raisebox{0pt}[12pt][6pt]{$32.8\pm7.6$} \\
\hline
\end{tabular}
\end{center}
\end{table}

   A clear signal has been found in the channel
$B^+ \to K^{*o}_X(1400) \pi^+$, $K^{*o}_X(1400) \to K^+\pi^-$. Among the resonances 
in this region, $K_1(1270)$,  $K_1(1400)$, $K^*(1410)$, $K^*_o(1430)$ and $K^*_2(1430)$,
the first two do not decay into $K\pi$ at all. The branching fraction of $K^*(1410) \to K\pi$
is too small to make its contribution essential. Thus, only two candidates remain:
$K^*_o(1430)$ and $K^*_2(1430)$. However, in the factorization approximation which 
seems adequate in this case, the local production of the tensor $K^*_2(1430)$ is
strictly zero~\cite{buras}. Therefore, its production is only possible 
due to corrections to the factorizable contribution and most probably is highly 
suppressed.

   In contrast, the scalar meson $K^*_o(1430)$ can be easily produced via the main
factorizable contribution and according to a recent estimate~\cite{chernyak}
the branching fraction for $B^+\to K^*_o(1430)\pi^+$ is even larger than that for
$B^+\to K^o\pi^+$. The prediction gives 
${\cal{B}}(B^+\to K^*_o(1430)\pi^+)\times{\cal{B}}(K^*_o(1430)\to K^+\pi^-) \simeq 28\times10^{-6}$
which is in good agreement with our result ($23.8^{+5.6}_{-5.1}\pm5.8)\times10^{-6}$.

   A significant signal was observed for the first time in the decay mode 
$B^+ \to f_o(980) K^+$ with a product of branching fraction of 
${\cal{B}}(B^+\to f_o(980)K^+)\times{\cal{B}}(f_o(980)\to \pi^+\pi^-) =
(16.4^{+4.9}_{-4.2}\pm 2.8)\times10^{-6}$ 
also indicates strong coupling of penguins with scalars. 

  It is not easy to identify a broad state with a mass around 1500 MeV
observed in the decay mode $B^+ \to K^+K^+K^-$. If an attempt
is made to describe the observed spectrum by a single resonance
in addition to the $\phi(1020)$, 
then it is natural to assume that it is a scalar $s\bar{s}$ state.
Such a state is hardly compatible with either one of the 
scalars $f_o(1370)$ and $f_o(1500)$  suggested by Particle 
Data Group~\cite{PDG} and having small probabilities of the decay into
$K\bar{K}$ final state.
Existence of a state with required properties is predicted in some
potential models~\cite{klempt} and evidence for such a state was reported by the 
LASS experiment~\cite{lass}. 

\begin{table}[t]
\begin{center}
\caption{Summary of the results on the search for $B^+\to~K^+h^+h^-$ decays.
\label{tab:result1}}
\vspace{0.2cm}
\begin{tabular}{|c|c|c|c|c|}  
\hline
\raisebox{0pt}[13pt][7pt]{Mode}     &
\raisebox{0pt}[13pt][7pt]{$\delta$} &
\raisebox{0pt}[13pt][7pt]{Yield}    &
\raisebox{0pt}[13pt][7pt]{\hspace*{-0.1cm}Sig., $\sigma$\hspace*{-0.1cm}}     &
\raisebox{0pt}[13pt][7pt]{${\cal{B}}(10^{-6})$} \\
\hline 
\raisebox{0pt}[13pt][7pt]{$K^+\pi^+\pi^-$} &
\raisebox{0pt}[13pt][7pt]{0.94}            &
\raisebox{0pt}[13pt][7pt]{\hspace*{-0.07cm}83.0$\pm$12.5}\hspace*{-0.07cm} &
\raisebox{0pt}[13pt][7pt]{8.6}             &
\raisebox{0pt}[13pt][7pt]{64.8$\pm$10.$\pm$7.0}\\
\hline 
\raisebox{0pt}[13pt][7pt]{\hspace*{-0.1cm}$K^+K^+K^-$\hspace*{-0.1cm}} &
\raisebox{0pt}[13pt][7pt]{0.72} &
\raisebox{0pt}[13pt][7pt]{$48.0\pm8.0$} &
\raisebox{0pt}[13pt][7pt]{9.3} &
\raisebox{0pt}[13pt][7pt]{36.5$\pm$6.1$\pm$5.5} \\
\hline
\raisebox{0pt}[13pt][7pt]{$K^-\pi^+\pi^+$} &
\raisebox{0pt}[13pt][7pt]{0.94} &
\raisebox{0pt}[13pt][7pt]{$7.8^{+6.2}_{-5.5}$} &
\raisebox{0pt}[13pt][7pt]{--} &
\raisebox{0pt}[13pt][7pt]{$<12.8$} \\
\hline
\raisebox{0pt}[13pt][7pt]{$K^+K^-\pi^+$} &
\raisebox{0pt}[13pt][7pt]{0.83} &
\raisebox{0pt}[13pt][7pt]{$9.7^{+6.6}_{-5.9}$} &
\raisebox{0pt}[13pt][7pt]{--} &
\raisebox{0pt}[13pt][7pt]{$<17.0$} \\
\hline
\raisebox{0pt}[13pt][7pt]{$K^+K^+\pi^-$} &
\raisebox{0pt}[13pt][7pt]{0.83} &
\raisebox{0pt}[13pt][7pt]{$0.0^{+2.6}_{-0.0}$} &
\raisebox{0pt}[13pt][7pt]{--} &
\raisebox{0pt}[13pt][7pt]{$<5.2$} \\
\hline
\end{tabular}
\end{center}
\end{table}

Similar uncertainties arise in the interpretation of the peak with a 
$\pi^+\pi^-$ mass  about 1300 MeV in the $K\pi\pi$ system.      
Two candidates for such a state - $f_2(1270)$ and $f_o(1370)$
exist. The assumption of the $f_o(1370)$ mechanism 
with its rather small coupling~\cite{anisov} to $\pi^+\pi^-$ 
would lead to an unnaturally large branching ratio of the $B$ decay.
The parameters of the bump obtained from the fit are close to those of
the $f_2(1270)$.
However, as recently shown~\cite{kim}, factorization leads to a very small
branching ratio for $B^+\to f_2(1270)K^+$.
If our observation is shown to be caused by the $f_2(1270)$, then
it provides evidence for the nonfactorized contributions 
which could also account for the large yield of $\eta'K$ in $B$ decays.

\begin{table}[t]
\begin{center}
\caption{Summary of the results on the search for intermediate resonances in
$B^+\to K^+\pi^+\pi^-$ and $B^+\to K^+K^+K^-$ states.
\label{tab:result2}}
\vspace{0.2cm}
\begin{tabular}{|c|c|c|c|c|}  
\hline
\raisebox{0pt}[15pt][7pt]{Mode}                       &
\raisebox{0pt}[15pt][7pt]{$\delta$}              &
\raisebox{0pt}[15pt][7pt]{Yield}                      &
\raisebox{0pt}[15pt][7pt]{\hspace*{-0.1cm}Sig., $\sigma$\hspace*{-0.1cm}}     &
\raisebox{0pt}[0pt][7pt]{${\cal{B}}_{Xh}\times{\cal{B}}_{hh}$} \\
 & & &  & \raisebox{0pt}[8pt][5pt]{$(10^{-6})$}  \\
\hline 
\raisebox{0pt}[13pt][7pt]{$K^{*o}(892)\pi^+$}         &
\raisebox{0pt}[13pt][7pt]{\hspace*{-0.1cm}0.92\hspace*{-0.1cm}}                   &
\raisebox{0pt}[13pt][7pt]{$7.5^{+4.1}_{-3.4}$}   &
\raisebox{0pt}[13pt][7pt]{2.8}                        &
\raisebox{0pt}[13pt][7pt]{$<11.6$} \\
\hline
\raisebox{0pt}[13pt][7pt]{\hspace*{-0.1cm}$K^{*o}_X(1400)\pi^+$\hspace*{-0.1cm}}      &
\raisebox{0pt}[13pt][7pt]{0.89}                   &
\raisebox{0pt}[13pt][7pt]{$29.1^{+6.9}_{-6.3}$}       &
\raisebox{0pt}[13pt][7pt]{5.2}                        &
\raisebox{0pt}[13pt][7pt]{$23.8^{+5.6}_{-5.1}$$\pm$5.8} \\
\hline
\raisebox{0pt}[13pt][7pt]{$\rho^o(770)K^+$}           &
\raisebox{0pt}[13pt][7pt]{0.67}                   &
\raisebox{0pt}[13pt][7pt]{$5.5^{+4.1}_{-3.4}$}        &
\raisebox{0pt}[13pt][7pt]{1.6}                        &
\raisebox{0pt}[13pt][7pt]{$<13.5$} \\
\hline
\raisebox{0pt}[13pt][7pt]{$f_o(980)K^+$}              &
\raisebox{0pt}[13pt][7pt]{0.79}                   &
\raisebox{0pt}[13pt][7pt]{$17.9^{+5.3}_{-4.6}$}       &
\raisebox{0pt}[13pt][7pt]{4.9}                        &
\raisebox{0pt}[13pt][7pt]{$16.4^{+4.9}_{-4.2}$$\pm$2.8} \\
\hline
\raisebox{0pt}[13pt][7pt]{\hspace*{-0.1cm}$f_X(1300)K^+$\hspace*{-0.1cm}}             &
\raisebox{0pt}[13pt][7pt]{0.79}                   &
\raisebox{0pt}[13pt][7pt]{$20.4^{+5.7}_{-5.0}$}       &
\raisebox{0pt}[13pt][7pt]{4.8}                        &
\raisebox{0pt}[13pt][7pt]{$18.8^{+5.3}_{-4.6}$$\pm$4.3} \\
\hline
\raisebox{0pt}[13pt][7pt]{\hspace*{-0.1cm}$f_X(1500)K^+$\hspace*{-0.1cm}}             &
\raisebox{0pt}[13pt][7pt]{0.45}                   &
\raisebox{0pt}[13pt][7pt]{$20.8^{+5.5}_{-4.8}$}       &
\raisebox{0pt}[13pt][7pt]{5.5}                        &
\raisebox{0pt}[13pt][7pt]{$25.1^{+6.6}_{-5.8}$$\pm$4.3} \\
\hline
\end{tabular}
\end{center}
\end{table}

Any unambiguous conclusion about the substructure of the observed signal
will require a larger data sample which will then allow us to perform
the angular analysis of the decay products of the intermediate resonances.

\end{document}